# Workshop on Combating Biases and Improving Institutional Culture

Deepa Chari[1, a)] and Vandana Nanal[2]

[1]*Homi Bhabha Centre for Science Education, Mumbai, India*
[2]*Tata Institute of Fundamental Research, Mumbai, India*

[a)] Corresponding author: deepa@hbcse.tifr.res.in

**Abstract.** The imposter syndrome, implicit biases, and microaggressions are some of the problems that adversely impact gender minorities in general. This workshop, conducted as a satellite event of the International Conference on Women in Physics 2023, is an attempt to create a platform for elaborating these concepts and sharing mutual experiences within the global Physics community. The larger goal is to develop a better understanding of the environments in Physics institutions worldwide. Through a series of activities and discussions, various exemplar scenarios were presented to participants and subsequent in-depth discussion focused on strategies to combat these issues as well as avenues for professional and academic support.

## INTRODUCTION

Physics continues to remain one of the least gender-balanced domains worldwide, with underrepresentation being more significant at the higher education levels. The scenario in India is broadly similar to that at the global level. Improving gender diversity in higher education in Physics is a topic of national interest and has been one of the focus areas of the Gender in Physics Working Group (GIPWG) of the Indian Physics Association.

It is known that the disproportionalities rooted in other sociocultural factors including race-ethnic representation, language, socio-economic status, etc., compound the gender-underrepresentation challenge. Studies have shown that underrepresentation is strongly correlated to an unfriendly/hostile environment, feeling of inadequacy, biases, etc. Moreover, people at different levels (students, teachers, researchers) may disproportionately experience some of these issues, which could not only leave enervating effects in their lives but may also affect their career choices. By connecting these threads, we emphasize that the problem of gender underrepresentation has its roots connected to the portrayal of the Physics subject to young learners, nature of the Physics community, and nature of society. Therefore, it is important to understand the experiences of the Physics community members in their contextual surroundings. The 'combating biases and improving institutional culture' workshop aims to explore how the participants relate to the three notions, namely, feeling of imposter, implicit bias, and microaggression, with their contextual surrounding. Information from their experiences and the environments of Physics departments, where they interact with the Physics community, is an essential input to develop strategies for overcoming some of these challenges.

The workshop design drew from sociocultural theories of self-identity, mindset theories, and theories of social dominance [1, 2]. It emphasized on generating collective discourse on challenges in career paths and providing opportunities to critically reflect on shared experiences as a way to learn. The goal was to open the discussions with the global community and learn about combating strategies at the individual, peer, and institutional levels along with participants.

It should be mentioned that this workshop was adapted from Program for Aspiring Women Scientists (PAWS) under the umbrella of GIPWG to reach out to the wider Physics community for discussions about combating challenges in a career in Physics [3]. This itself was seeded from an interactive workshop designed for an annual three-week summer program called Vigyan Vidushi (Physics), for women students in India pursuing their Master's degree in Physics [4]. The PAWS workshops were hosted mostly in the online mode by the Physics departments of various institutes in India and authors interacted with more than 200 Physics/Engineering undergraduate, postgraduate, and graduate (Ph.D.) students in India. The response has been extremely positive and has led to a greater repertoire of experiential examples/scenarios on implicit bias, feeling of imposters, and microaggressions. Further, anecdotal evidences suggest the need for continuous engagement with these issues with students, especially at the important junctures of their career decisions.

## NATURE OF THE WORKSHOP

The workshop was conducted on the Zoom platform for a duration of two hours. A total of 42 participants spanning different regions of Asia, Africa, Europe, and America (north and south) attended the workshop. The participant group was a mixture of graduate students and early-career professionals including faculty and scientists associated with Physics. The workshop was open to female participants only.

At the start of the workshop, participants were briefed about the general workshop guidelines and norms, and were requested keep pen-paper or an equivalent device for noting down observations. The workshop started with a theoretical introduction of the notions of imposter (self-complex), implicit bias, and microaggression, which was followed by two activities: 1) 'feeling of imposter' and 2) 'observing bias and microaggressions'. The participants were divided in smaller groups for discussions at various points during the session.

The activity on the feeling of imposter involved a description of seven different plausible scenarios such as applying for a summer internship, Ph.D. application, presenting a research paper, discussions group meeting, reviewing student/teacher performance in the Physics course, etc. Each scenario was tagged with some emotional experience such as feeling lucky, anxious, under-deserving, favored, etc. During the workshop, the facilitator displayed one scenario at a time and requested each participant to assign a numerical rank to each scenario on a scale of +2 (strongly agree) to –2 (strongly do not agree). Participants were expected to relate their feelings about the situation from their past experience and rank the scenario individually. An example is shown in Fig. 1(a) below.

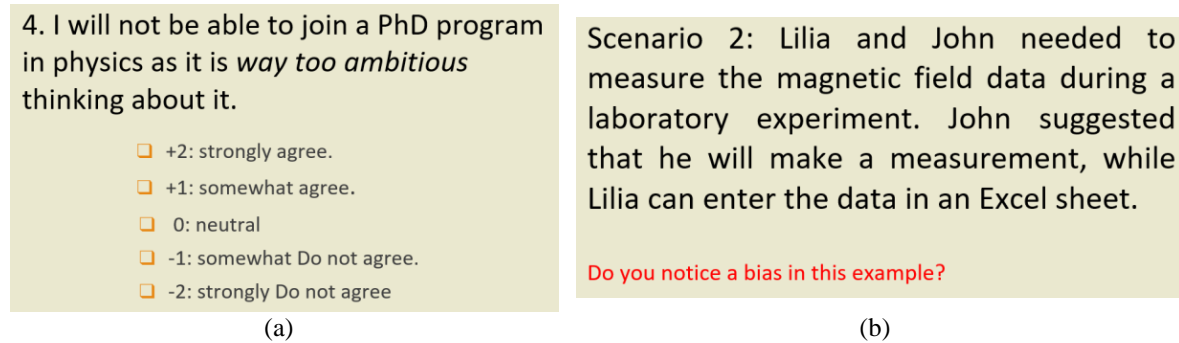

(a)                                                            (b)

**FIGURE 1:** Example of a scenario from (a) activity 1 on imposter syndrome and (b) activity 2 related to biases.

Ample time was provided to the participants to think and rank each scenario. At the end, they were requested to report the sum total of the numerical ranks and this was shared with all participants. Participants were encouraged to reflect on the emerged pattern of numbers, and discussion highlighted how this, in turn, connected to individual and collective feelings about the imposter phenomenon. The activity was conducted as one large group, and later, the participants were divided into three subgroups (using the breakout room option in Zoom) to continue the extended discussions.

In the second activity, participants were divided into smaller groups (~five members each) and one member was designated as a coordinator. Each group was presented with a unique situation that involved either a sense of bias, self-complex or microaggression, or none. Each group deliberated on the situation presented to them and their interpretation about the same. A typical example of a situation presented for discussion is shown in Fig. 1(b). Subsequently, all participants met together as a group where the group coordinator presented a summary of their deliberation. Further discussion in the group provided useful insights into understanding various concepts and possible ways to deal with each example.

In addition to the prescribed activities, additional opportunities were also facilitated for participants to interact with each other and share their pertinent experiences. At the end of the workshop, feedback was collected about the activity content, pace, and utility for constructive revisions in future.

## OUTCOMES AND REFLECTION

During the first activity on the imposter syndrome, the participants discussed how they could closely associate with many of the presented scenarios. The observed randomness of the sum pattern emphasized the contextual nature of the feeling of imposter. In other words, one might not feel the same way about every situation. It was noted that generally one was good at something but might be fearful about some other things, and the sense of imposter might evolve with time. It was emphasized that the discomfort about a new course/new research topic/any situation could be natural. It is important to realize this feeling of imposter and find ways to deal with it.

A positive message "not to undermine one's own strength" emerged from the workshop activity. Further, suggestions regarding setting up of a support network with the help of student groups/advisors/friends to cope with situations involving the imposter feeling in academia were presented.

The second activity on identifying biases, microaggression, and imposter situations had a mixed response. Different groups shared their observations and responses about the scenarios. Various strategies of combating biases were discussed in depth in small groups. It was interesting to engage in discussions with participants from diverse backgrounds.

Activity-specific feedback was collected by requesting the participants to rate each of the activities on a scale of 0 (poor) to 5 (excellent). Approximately 50 % of the participants volunteered and provided feedback, which is shown in the Figure 2.

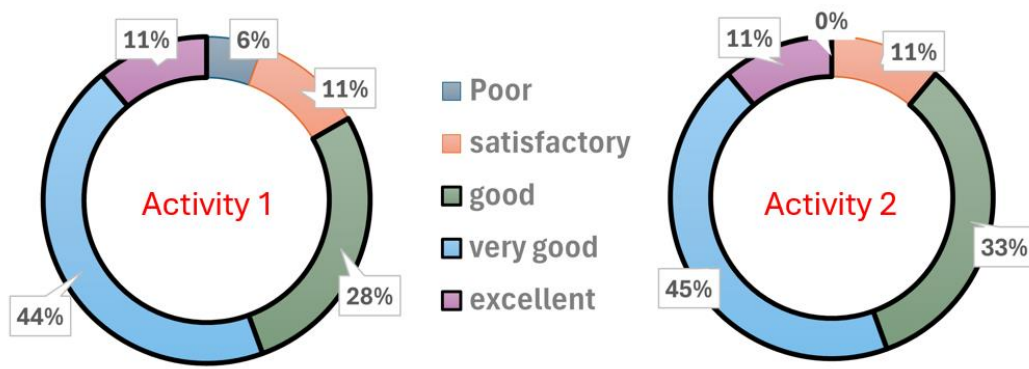

**FIGURE 2:** Ratings of the workshop activities.

All participants informed that they found the workshop very useful and would recommend it to their friends. The overwhelming positive response (> 80 % ranking it good to excellent for both activities) is a measure of the success of this workshop. The positive feedback from diverse groups of participants reaffirms the belief that the design of the workshop allowed participants to be more agentic in terms of reviewing/analyzing the scenarios within their contextual environments and consequently their strategies had greater authenticity than any readymade/fixed solutions for combating biases.

It was also evident that global platforms such as the ICWIP conference presented a safe place, where the concerned Physics community could freely share their perspectives and experiences on various aspects of underrepresentation in Physics. Further, the online platform facilitates a wider reach and the workshop series will be continued in future in online and other possible modalities.